\documentclass[%
 reprint,
showpacs,
 amsmath,amssymb,
 aps,
]{revtex4-1}

\usepackage{graphicx}
\usepackage{dcolumn}

\usepackage{graphicx}
\usepackage{dcolumn}
\usepackage{bm}

\begin{document}

\title{Resolution Limit of Label-free Far-field Microscopy.}



\author{Evgenii E. Narimanov}
\affiliation{School of Electrical and Computer Engineering  and Birck Nanotechnology 
Center, \\ Purdue University, West Lafayette, IN 47907, USA
}%


\begin{abstract}
The Abbe's diffraction limit\cite{Lagrange,Helmholtz,Abbe} that relates the maximum optical resolution to the numerical aperture of the lenses involved and the optical wavelength, is generally considered as ``a practical frontier that cannot be overcome with a conventional imaging system.'' \cite{Goodman_book} However, it does not represent a fundamental limit to the optical resolution, as demonstrated with several new imaging techniques that proved the possibility of finding the subwavelength information from the far-field of an optical image, from super-resolution fluorescence microscopy \cite{Betzig,Hell} to the imaging systems that use new data processing algorithms leading to a dramatically  improved resolution \cite{SegevSparse1,SegevSparse2,SegevSparse3} to super-oscillating  metamaterial lenses.\cite{ZheludevSO1,ZheludevSO2,ZheludevSO3} This raises the key question of whether there's in fact a fundamental bound to the optical resolution -- as opposed to ``practical'' limitations due to noise and imperfections, and if so then what it is. In the present work, we derive the fundamental limit to the resolution of optical imaging, and demonstrate that, while a bound to the resolution of a fundamental nature does exit, contrary to the conventional wisdom it is neither exactly equal to nor necessarily close to Abbe's estimate. Both the  exact  value of the fundamental resolution limit uncovered in the present work, and the physical mechanism behind it, would help the development of imaging systems that offer the optimal performance in the practical environment with the often contradictory requirements for imaging resolution, speed and robustness to noise. Furthermore, our  approach  to imaging resolution that combines the tools from the physics of wave phenomena and the methods of information theory, is general, and can  be extended beyond optical microscopy,  to e.g.  geophysical and ultrasound imaging.
\end{abstract}

\maketitle

High resolution optical imaging holds the key to the understanding of fundamental microscopic processes both in nature and in artificial systems -- from the charge carrier dynamics  in electronic nano-circuits \cite{electrons_imaging} to the biological activity in cellular structures.\cite{biological_optics}  However, optical diffraction that prevents the ``squeezing'' of light into the dimensions much smaller than its wavelength,  
does not allow a straightforward extension of the conventional optical microscopy to the direct imaging of such subwavelength structures as cell membranes, individual viruses or large protein molecules. As a result, recent decades have seen increasing interest in developing ``super-resolution'' optical methods that allow to overcome this diffraction barrier -- from near-field optical microscopy \cite{NSOM} to structured illumination imaging \cite{SIM} to metamaterials-based super-resolution \cite{metaSR} to two-photon luminescence and stimulated emission depletion microscopy  \cite{Hell} to   stochastic optical reconstruction imaging \cite{STORM} and photoactivated localization microscopy. \cite{Betzig} 

In particular, there is an increasing demand for the approach to optical imaging that is inherently label-free and does not rely on fluorescence, operates on the sample that is in the far-field from all elements of the imaging system, and offers the resolution comparable to that of the fluorescent microscopy. While seemingly a tall order, this task has recently found two possible solutions -- that approach the problem from the ``hardware'' and ``algorithmic'' sides respectively. The former approach relies on the phenomenon of ``super-oscillations'' --  where the band-limited function can, and -- when properly designed --  does oscillates faster that its fastest Fourier component. The actual super-oscillatory lenses that implement this behavior, have been designed and fabricated,\cite{ZheludevSO2,ZheludevSO3} and optical resolution exceeding the conventional Abbe's limit, has been demonstrated in experiment.\cite{ZheludevSO2}  The second approach relies on new methods of processing the ``diffraction-limited'' data, taking full advantage of the fact that actual targets (and especially biological samples) are often inherently sparse.\cite{biological_optics} The resulting resolution improvement beyond the Abbe's limit due to this improved data processing has been demonstrated both in numerical simulations and in experiment.\cite{SegevSparse1,SegevSparse2,SegevSparse3}

Far-field optical resolution beyond the Abbe's limit in a scattering, rather than fluorescence - based approach, observed in Refs. \cite{SegevSparse1,SegevSparse2,SegevSparse3,ZheludevSO1,ZheludevSO2,ZheludevSO3}, clearly demonstrates that Abbe's bound of half-wavelength (and its quarter-wavelength counterpart for structured illumination) is not a fundamental limit for optical imaging. This raises the key question of whether there's in fact a fundamental bound to the optical resolution -- as opposed to ``practical'' limitations due to detector noise, imaging system imperfections, data processing time limits in the case when image reconstruction corresponds to an NP-complete problem, etc. Furthermore, the knowledge of the corresponding fundamental limit, if such exists, and the physical mechanism behind it, would help finding the way to the system that offers the optimal performance --  just as deeper understanding of thermodynamics and Carnot's limit helped the design of practical heat engines. 

In the present work, we show that there is in fact a fundamental  limit on the resolution of far-field optical imaging, which is however much less stringent than Abbe's criterion. 
 The presence of any finite amount of noise in the system, { regardless of how small  is its intensity},  leads to a fundamental limit on the optical resolution, that can be expressed in the form of an effective uncertainly relation. This limit has essential information-theoretical nature, and can be connected to the Shannon's theory of information transmission in linear systems.\cite{Shannon}

\subsection*{Definition of the resolution limit}

We define the diffraction limit $\Delta$ as the shortest spatial scale of the object whose geometry can still be reconstructed, {\it error-free}, from the far-field optical measurements in the presence of noise.\cite{footnote} Without loss of generality, one can then assume that the object is composed of arbitrary number of point scatterers of arbitrary amplitudes located at the nodes of the grid with the period $\Delta$, as any additional structure in the sources (or scatterers \cite{footnote2}) or variations in position, will add to the information that needs to be recovered from far-field measurement for the successful reconstruction of the geometry of the object. 

Furthermore, the essential ``lower bound'' nature of $\Delta$ further allows to reduce the problem to that of an effectively one-dimensional target (formed by line, rather than point, sources) -- since, as it was already known to M. Andr\'e \cite{Andre1876} and L. Raleigh,\cite{Raleigh1879}  line sources are ``more easily resolvable'' than point sources. 

\begin{figure}[htbp] 
   \includegraphics[width=3.5in]{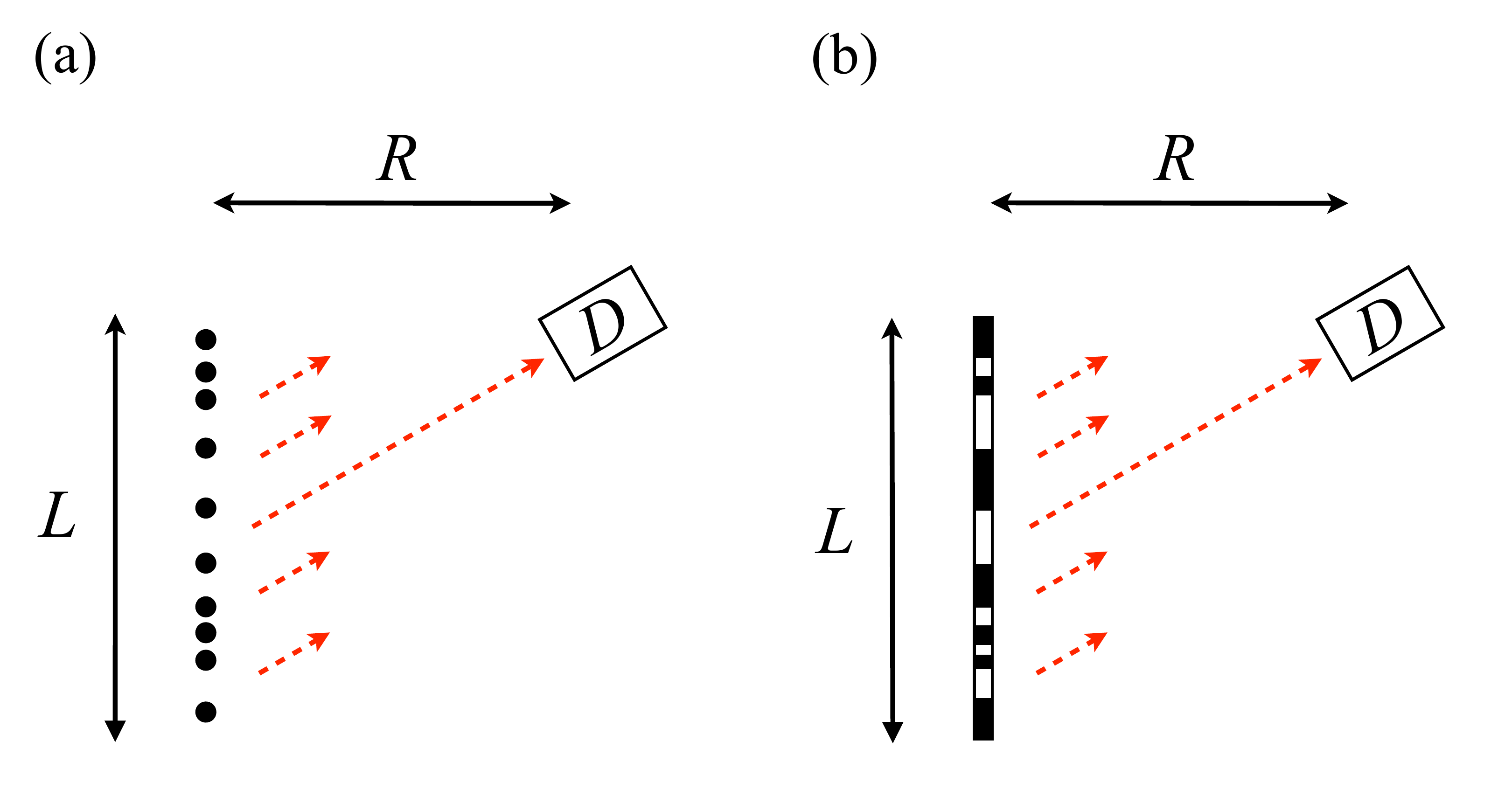} 
   \caption{ The schematic representation of the imaging set-up, for the object formed by an array of small 
particles / lines (a) and a (binary) mask (b). $D$ labels the position of a (coherent) detector, $L$ is the size of the object (and equivalently the imaging aperture), and $R$ is the distance from the object to the detector; in the far field 
$R \gg L$.  }
   \label{fig:schematics}
\end{figure}

To calculate the fundamental resolution limit, is is therefore sufficient to consider the model system of an array of line ``sources'' of arbitrary (including zero) amplitudes, located at the node points of the grid with the period $\Delta$ -- see Fig. \ref{fig:schematics}(a). Note that, in term of the information that is detected in the far field and the information that is necessary and sufficient for the target reconstruction, this problem is identical to that of a step mask where thickness and/or permittivity changes at the nodes of the same grid by the amounts proportional to the amplitudes of the corresponding line sources (as the point source distribution corresponds to the spatial derivative of the mask ``profile'') -- 
see Fig. \ref{fig:schematics}(b).

Note that the reduction of the original problem to that of an effectively one-dimensional profile is not a simplification for the sake of convenience or reduction of the mathematical complexity of the problem. It is exactly this ``digitized'' one-dimensional profile that corresponds to the smallest 
``resolvable'' spatial scale among all objects with a low bound on their spatial variations, and therefore defines the fundamental resolution limit. Furthermore, in many cases the actual object is formed by two (or more) materials that form sharp interfaces. In this case, the step mask that is equivalent to our point source model, offers an adequate representation of the actual target.

However, even within the original framework of ``resolving'' two point sources,\cite{Raleigh1879} the result clearly depends on the difference of their amplitudes -- with increasing disparity between the two leading to progressively worse ``resolution''. The ``ultimate'' resolution limit $\Delta$ therefore corresponds to the case of identical point sources (or subwavelength scatterers), which are present only in an (unknown) fraction of the  grid nodes.  Note that such digital mask corresponds to the common case of a pattern formed by a single material (and e.g. the surrounding air) -- see Fig. \ref{fig:schematics}(b).

When the distance to the detector $R$ is much larger than the aperture $L$, $R \gg L$ (see Fig. \ref{fig:schematics}), for each polarization we find (see Appendix A)
\begin{eqnarray}
s\left({\bf k}\right)  & = & \sum_i  \alpha_i E_0\left({\bf \boldsymbol\rho}_i\right) \exp\left(i {\bf k}\cdot \boldsymbol\rho_i\right) + n\left({\bf k}\right),
\label{eq:se1}
\end{eqnarray}
where ${E}_0$ is the incident field ``illuminating'' the target, $i$ is the (integer) index that labels the (point) scatterers with the corresponding polarizabilities $\alpha_i$, ${\bf \boldsymbol\rho}_i \equiv \left(x_i, y_i\right)$,  ${\bf k} \equiv \left(k_x, k_y\right)$ is the wavevector with the magnitude $\left| {\bf k} \right| < \omega/c \equiv k_0$, $\omega$ is the light frequency, and $c$ is the speed of light (in the medium surrounding the target).
Equivalently, for the case of the object in the form of a (dielectric) mask (see Fig. \ref{fig:schematics}(b)), we obtain
\begin{eqnarray}
s\left({\bf k}\right)  & = & \int d^2{\bf \boldsymbol\rho} \ \Delta\epsilon\left({\bf \boldsymbol\rho}\right)  \ E_0\left({\bf \boldsymbol\rho}_i\right) \exp\left(i {\bf k}\cdot \boldsymbol\rho_i\right) + n\left({\bf k}\right), \ \ \ \ \ 
\label{eq:se2}
\end{eqnarray}
where $\Delta\epsilon$ is the difference between the dielectric permittivities of the object and the background.

Here $n\left({\bf k}\right)$ corresponds to the effective noise which includes the contributions from all origins (detector dark currents, illumination field fluctuations, etc.). Using data for imaging with different electromagnetic field polarizations, the effective noise can be correspondingly reduced.

Note that the model of Eqn.  (\ref{eq:se1}) or its equivalent (\ref{eq:se2}) assumes coherent detection of the electromagnetic field in the far zone. This is essential for the definition of the fundamental resolution limit, as the phase information is in fact available in the far field, and can be measured even with an intensity only sensitive detector using e.g. optical heterodyne approach,\cite{heterodyne} so that any failure to obtain the corresponding information in a given experimental setup cannot be attributed to the fundamental resolution limit of optical imaging.

Finally, for the calculation of the fundamental resolution limit $\Delta$ we must assume the large aperture limit $k_0 L \gg 1$. While the case of a small aperture $k_0 L \leq 1$ can be easily implemented in the actual experimental setup (albeit at the cost of dramatic reduction in the field of view), 
the aperture in a close proximity to the object represents an example of a near-field probe, and this setup cannot be treated as a true far-field imaging.

\subsection*{Information-theoretical framework}

To derive the fundamental  limit on the  resolution of optical imaging, we calculate the total amount of information about the object that can be recovered in the far field. As our Eqn. (\ref{eq:se1}) can be interpreted as the input ($E_0$) -- output ( $\left\{s\right\}$ ) relation of a linear information channel, the amount of the actual information carried from the object to the far field detector, can be calculated using the standard methods of the information theory.\cite{Shannon}
The resolution limit then follows from the requirement of the recovered information being sufficient to reconstruct the target:
\begin{eqnarray}
\Delta & = & \frac{L}{T}. 
\label{eq:delta1}
\end{eqnarray}

When the object is composed of $M$ different materials (or is formed by an array of point sources with $M$ different
levels of amplitude), additional information is needed for its reconstruction, which leads to a more stringent bound on the spatial resolution,
\begin{eqnarray}
\Delta_M & = & \frac{L  \log_2M }{T } = {\Delta}\cdot{\log_2M} \geq \Delta.
\label{eq:deltaM}
\end{eqnarray}

The actual transmitted information $T$ can be obtained from the mutual information functional \cite{Shannon}
\begin{eqnarray}
T & = & H\left[\left\{ s\right\}\right] - H\left[\left\{ s\right\} \left|\right. E_0, \alpha\right].
\label{eq:T}
\end{eqnarray}

Here, the entropy $H\left[\left\{ s\right\}\right]$ is the measure of the information received at the detector array. 
\begin{eqnarray}
H\left[\left\{ s\right\}\right] & = & - \int {\cal D}s\left({\boldsymbol k}\right) \ P\left[ s\left({\boldsymbol k}\right)\right] 
\ \log_2P\left[ s\left({\boldsymbol k}\right)\right]. 
\label{eq:unconditional}
\end{eqnarray}

However, as the system is noisy, for any output signal, there is some uncertainty of what was the originating field at the mask. The conditional entropy $H\left[\left\{ s\right\} \left|\right. E_0\right]$ at the detector array {\it for a given} $E_0$ represents this uncertainty:
\begin{eqnarray}
H\left[\left\{ s\right\}\left|\right. E_0 \right] & = & - \int {\cal D}s\left({\boldsymbol k}\right) \ P\left[ s\left({\boldsymbol k}\right)\left|\right. E_0\left({\boldsymbol\rho}\right) \right] \nonumber \\
& \times &  \log_2P\left[ s\left({\boldsymbol k}\right)\left|\right. E_0\left({\boldsymbol\rho}\right),\alpha\right]. 
\label{eq:H_conditonal}
\end{eqnarray}

Substituting the resulting analytical expressions for $H\left[\left\{ s\right\}\right]$ and $H\left[\left\{ s\right\} \left|\right. E_0\right]$ (see {Appendix C}) into the mutual information $T$ in Eqn. (\ref{eq:T}), for the resolution limit in the case
of uniform illumination (see Appendix D for resolution limit in the regime of structured illumination) we obtain 
\begin{eqnarray}
\Delta & = &  \frac{\lambda}{2} \ \frac{1}{\log_2\sqrt{ 1 + 2 \ {\rm SNR} + \eta \ {\rm SNR}^2  }+ {\cal{O}}\left(1/k_0 L\right)}.
\ \ \ 
\label{eq:delta}
\end{eqnarray}
Here ${\rm SNR}$ is the effective signal-to-noise ratio measured at the detector array
\begin{eqnarray}
{\rm SNR} & = & \frac{\langle \left| s\left({\boldsymbol k}\right) - \langle s\left({\boldsymbol k}\right) \rangle\right|^2 \rangle}{\langle \left| n\left({\boldsymbol k}\right)  \right|^2 \rangle},
\label{eq:SNR}
\end{eqnarray}
and
\begin{eqnarray}
\eta & = & \frac{\langle \left({\rm Re}\left[ \alpha - \langle \alpha \rangle\right]\right)^2\rangle \langle  \left({\rm Im}\left[ \alpha - \langle \alpha \rangle\right]\right)^2 \rangle}{\langle \left| \alpha - \langle \alpha \rangle \right|^2\rangle} 
 \nonumber \\
 & - & 
 \frac{\langle {\rm Re}\left[ \alpha - \langle \alpha \rangle\right] 
 {\rm Im}\left[ \alpha - \langle \alpha \rangle\right] \rangle^2}{\langle \left| \alpha - \langle \alpha \rangle \right|^2\rangle}
 \label{eq:eta}
\end{eqnarray}
represents the relative contribution of the absorption in the target; for a transparent object (${\rm Im}\left[\alpha\right] = 0$) we have $\eta=0$. The correction ${\cal O}\left(1/k_0 L \right)$ accounts for the finite size of the imaging aperture, and can be neglected for $k_0 L \gg 1$.

\begin{figure}[htbp] 
   \includegraphics[width=3.in]{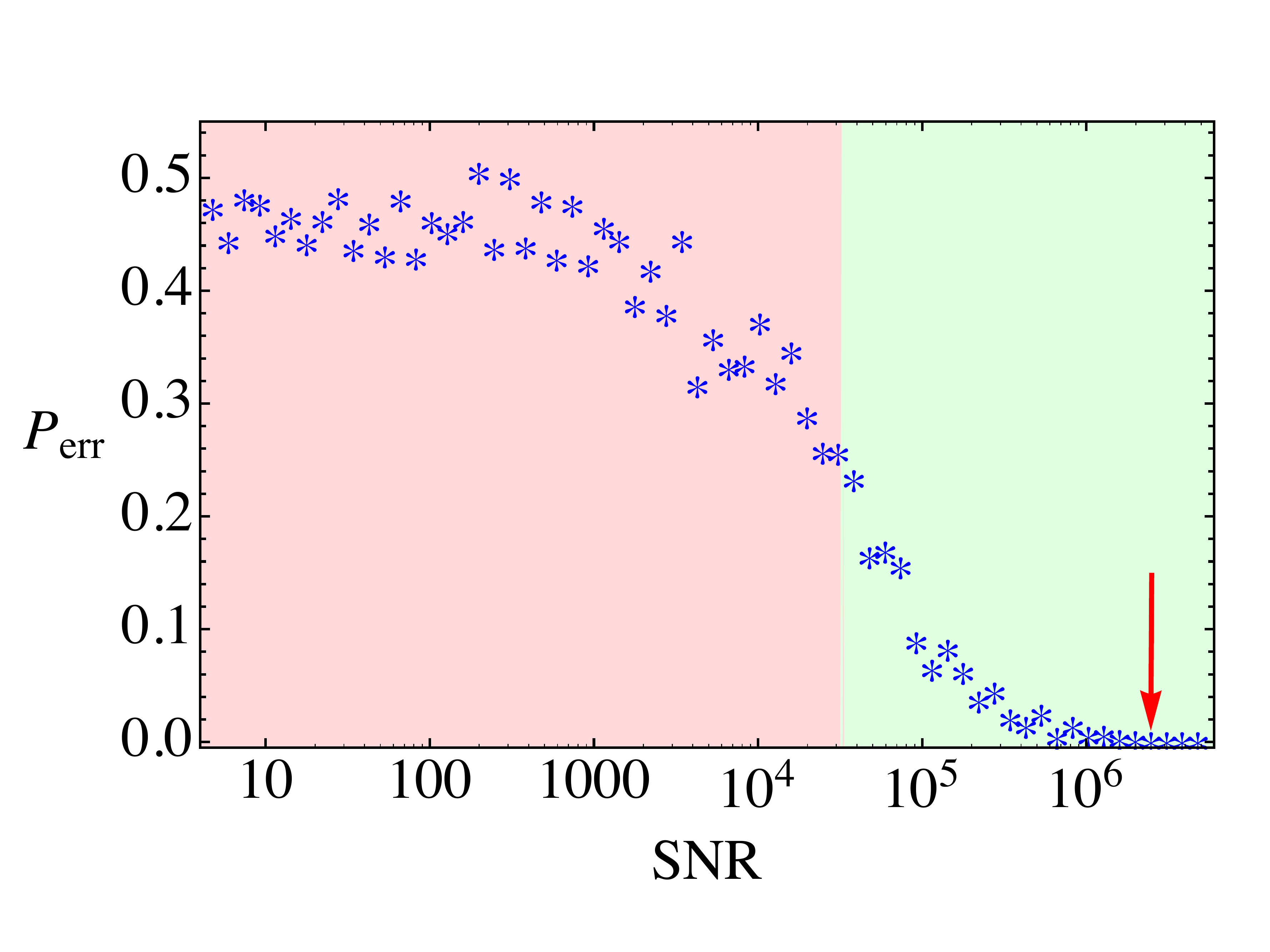} 
   \caption{ Super-resolution object reconstruction for a binary mask. The inset shows the schematics of the object profile.
   The main panel plots the error probability in the recovered profile, as a function of the effective signal-to-noise ratio, 
   ${\rm SNR}$. The data shown was obtained for 10000 different realizations. The boundary separating the light-red and light-green background, corresponds to the value of the signal-to-noise ratio corresponding to $\Delta$ sufficient to resolve the $\lambda/16$ spacing (see panel (a)). The red arrow indicates the minimum value of    
${\rm SNR}$ when the numerical reconstruction produces no errors.
   }
   \label{fig:simulation1}
\end{figure}

\subsection*{The Discussion}

While Eqn. (\ref{eq:delta}) allows for an unlimited resolution in a noise-free environment,  even a relatively low noise dramatically alters this picture. With the weak logarithmic dependence of the resolution limit on the ${\rm SNR}$,
to reduce the resolution limit by e.g. a factor of ten, the signal to noise ratio  needs to be increased by six orders of magnitude.

At the same time, the spatial resolution limit $\Delta_M$ depends on the effective ``uncertainty'' in the range of permittivity variations in the object that is being imaged -- the simpler is the structure of the target, the easier is the task of finding its geometry. The ultimate value $\Delta$ is then achieved in the case of a binary mask
(i.e. the object that is formed by only two materials), and represents the fundamental bound to the resolution. In the case of a higher complexity in the composition of the target, the actual resolution limit $\Delta_M$ is well above $\Delta$. When the number of materials (with their corresponding permittivities) that the object is composed of, $M$, is known {\it a priori}, the corresponding resolution limit $\Delta_M$ is defined by Eqn. (\ref{eq:deltaM}). However, when no {\it a priori} information whatsoever is available, the limit to the resolution can be expressed as the effective uncertainty relation, that offers the lower  bound on the 
product of the scaled spatial resolution $\delta_x$ and the amplitude resolution $\delta_\epsilon$. In the case when the object is composed of transparent materials (${\rm Im}\left[ \epsilon\right] = 0$) we obtain
\begin{eqnarray}
\delta_x \cdot F\left( \delta_\epsilon \right)\geq \frac{1}{\log_2\sqrt{1 + 2 \ {\rm SNR}} },
\label{eq:uncertainty}
\end{eqnarray}
where $F\left(t\right) \equiv 1/\log_2\left(1/t\right),$
the scaled spatial resolution $\delta_x$ is defined as the ratio of $\Delta$ to the Abbe's limit, 
and the scaled amplitude resolution corresponds to the uncertainty in the permittivity $\delta\epsilon$  that is normalized to the difference between the smallest ($\epsilon_{\rm min}$) and largest  ($\epsilon_{\rm max}$)  permittivities in the object,
$
\delta_\epsilon \equiv  {\delta\epsilon}/{\left(\epsilon_{\rm max} - \epsilon_{\rm min}\right)}.
$
For a binary mask,  $\delta\epsilon = \left(\epsilon_{\rm max} - \epsilon_{\rm min}\right)/2$, so that the scaled amplitude resolution $\delta_\epsilon = 1/2$, and $F\left(\delta_\epsilon\right) = 1$, which reduces the uncertainly relation (\ref{eq:uncertainty}) to the fundamental limit $\Delta$ of Eqn. (\ref{eq:delta}).

For imaging with no a priori information, with the optimal data reconstruction algorithm Eqn.  (\ref{eq:uncertainty}) represents a trade-off between the uncertainties in position and the amplitude of the recovered image.  Note that, as follows from Eqn. (\ref{eq:uncertainty}), spatial resolution at the Abbe's limit corresponds to the relative amplitude uncertainty of
at least $1/\sqrt{1 + 2 \ {\rm SNR}}$. 

In the case of imaging a binary mask or a pattern of identical subwavelength particles, the actual resolution can reach the value of $\Delta$ -- which for a high signal-to-noise ratio can be substantially below the Abbe's limit. For example, in the structured illumination setup with ${\rm SNR} \sim 10^{-6}$, we find $\Delta \sim \lambda/100$. While reaching all the way to this limit with the data obtained in the standard imaging setup may be highly nontrivial, a straightforward algorithm described below that implements the amplitude constraint, offers spatial resolution well below the Abbe limit
-- see Fig. \ref{fig:simulation1}. 

In the algorithm whose performance is shown in Fig. \ref{fig:simulation1}, the subwavelength binary mask 
(see the inset to Fig. \ref{fig:simulation1}) is recovered from its (band-limited) Fourier spectrum measured in the far-field, together with the constraint that limits its profile to only two values. While a finite amount of noise in the far-field measurements inevitably leads to errors, with the increase of the effective signal-to-noise ratio ${\rm SNR}$  the corresponding error probability $P_{\rm err}$ rapidly goes to zero. In particular,  for the resolution of $\lambda/16$ in the example of  Fig. \ref{fig:simulation1}, for the ${\rm SNR}$ beyond the value indicated by the red arrow, the numerical calculation with an ensemble of 10000 different realizations, showed no errors. 

The light-red and light-green color backgrounds in Fig. \ref{fig:simulation1}  corresponds to the parameter range that respectively violates and satisfies the fundamental resolution limit of Eqn. (\ref{eq:delta}). Note that the boundary separating these regimes, corresponds to the signal-to-noise ratio that is substantially less than the smallest value (shown by red arrow in Fig. \ref{fig:simulation1}) for the error-free performance in the data recovery -- indicating that the reconstruction algorithm is far from optimal. Still, even with this performance, the example of Fig. \ref{fig:simulation1}  indicates that even a straightforward implementation of an {\it a priori} constraint on the object geometry (binary mask rather than an arbitrary profile), offers object reconstruction from diffraction-limited data with deeply subwavelength resolution (four times below Abbe's limit in the example of Fig. \ref{fig:simulation1}).

Additional {\it a priori} information about the object further reduces the resolution limit of optical imaging. For different cases of a priori available information about the target, particularly 
important is the case of sparse objects, as this property is wide spread in both natural and artificial systems.\cite{biological_optics} If the
target is a priori known to be sparse, with the effective sparsity parameter $\beta$ (which can be defined as the fraction of empty ``slots'' in the grid superimposed on the target), we find
\begin{eqnarray}
\Delta^{(\beta)} & = & \frac{\Delta}{\left| \beta \log_2\beta + \left(1 - \beta\right)\log_2\left(1 - \beta\right)\right|}. 
\label{eq:delta_beta}
\end{eqnarray}
For the numerical example studied in Ref. \cite{SegevSparse1}, with $\beta \simeq 0.03$ and ${\rm SNR} \sim 10^2$,
the resolution limit $\Delta^{(\beta)} \simeq 0.025 \lambda$. Accurate  numerical reconstruction of the features on the scale of $\sim \lambda/10$ demonstrated in Ref. \cite{SegevSparse1}, is therefore fully consistent with the fundamental limit  $\Delta^{(\beta)}$.

\begin{figure*}[htbp] 
   \includegraphics[width=6.5in]{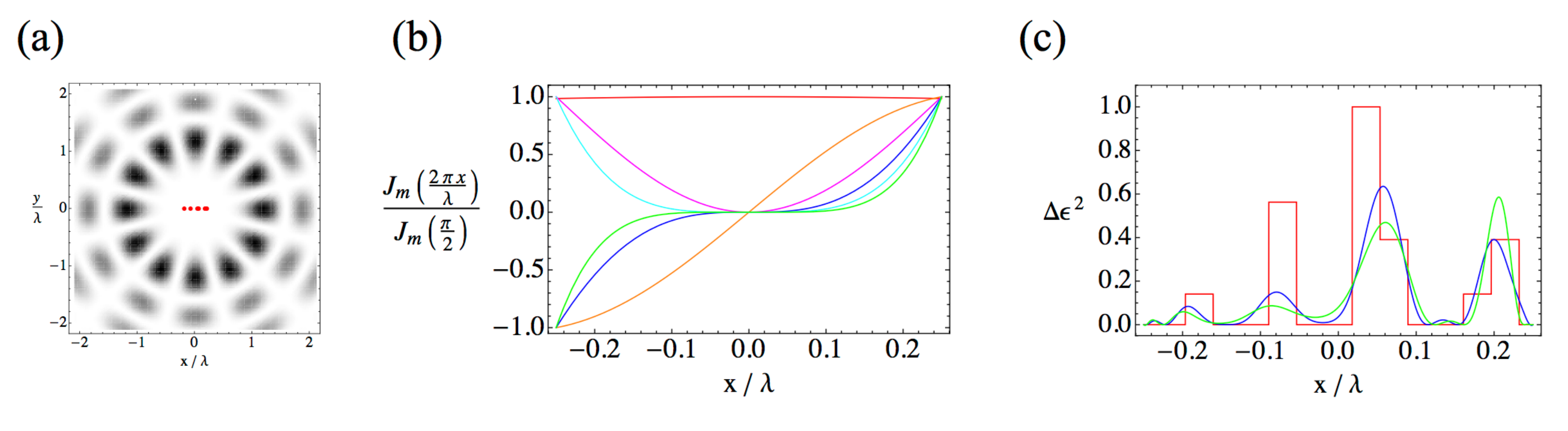} 
   \caption{ 
   Super-resolution imaging of a subwavelength object,  based on structured illumination with Bessel beams. Panel (a) plots the ``incident'' Bessel beam  of the order $m=12$ 
   (shown in gray-scale) focused at the center of the subwavelength object (red). Panel (b) shows the Bessel beam profiles in the object plane, for different orders $m =  0$ (red),  $1$ (orange), $2$ magenta), $3$ (blue), $4$ (cyan) and $5$ (green). For a small distance from the center, the Bessel function of order $m$ behaves as $x^m$, so illumination with the Bessel beams of different order effectively ``projects'' the target on the set $\{ x^m\}$ for different values of $m$. As the latter form a complete basis set, this procedure allows high-resolution reconstruction of the original object profile, without any use of super-oscillations or subwavelength focusing.  Panel (c) show  the  subwavelength object profile, and its reconstruction with Bessel beam illumination. The object corresponds to the red line in (c). The reconstructed profiles are shown for the effective signal-to-noise ratios of $10^6$ (blue line in panel (c)) and $10^4$ (green line in (c)). }
   \label{fig:simulation2}
\end{figure*}

\subsection*{Imaging with a small aperture}

The explicit expression for the resolution limit in Eqn. (\ref{eq:delta}) is  presented  in the large amplitude limit $k_0 L \gg 1$. While this corresponds to the most common regime of actual optical microscopy, using a small aperture that's comparable to the free space wavelength can offer its own advantages. The resulting effect on the resolution limit is accounted for by the (positive definite) term ${\cal O}\left(1/k_0L\right)$ in Eqn. (\ref{eq:delta}), which further reduces 
$\Delta$. 

Note that this was precisely the regime where super-oscillation based imaging was demonstrated in experiment, as the use of small aperture was essential to block the (exponentially) strong power side-lobes. While the resulting improvement of the resolution is consistent with the fundamental limit established in the present work, our expressions
(\ref{eq:deltaM}),(\ref{eq:delta}),(\ref{eq:delta_beta}) do not explicitly indicate the advantage of super-oscillations approach. This should be contrasted to the case of sparsity-based imaging where its key parameter $\beta$ explicitly enters the resolution limit in (\ref{eq:delta_beta}). 

Indeed, while the super-oscillations imaging does offer subwavelength resolution -- this improvement is the general feature of all structured illumination methods optimized for small aperture (or equivalently for imaging small isolated  objects), and is not limited to the super-oscillation approach. This behavior is illustrated in Figs. \ref{fig:simulation2}, where a subwavelength target (red pattern in the center of Fig. \ref{fig:simulation2}(a)) is illuminated by the Bessel beam propagating in the direction normal to the plane of the picture. The beam axis is ``focused'' to the center of  the target (see Fig. \ref{fig:simulation2}(a)), so that the illuminating field within the aperture not only shows no super-oscillations but in fact does not oscillate at all -- see the field profiles for different orders $m$ of the illuminating Bessel beams in Fig.  \ref{fig:simulation2}(b)). Nevertheless, the standard data recovery algorithm clearly shows deep subwavelength resolution of $\sim \lambda/10$ -- see Fig.  \ref{fig:simulation2}(c), despite having no {a priori} information about the structure of the target.

It should however be noted, that the super-oscillations based approach, when implemented to form a subwavelength focus spot that's used to scan the object,\cite{ZheludevSO2,ZheludevSO3} is naturally suitable for optical imaging limited to incoherent detection, which offers substantial practical advantages in the actual implementation of the system.

\subsection*{Conclusions}

In conclusion, we have derived the fundamental resolution limit for  far-field optical imaging, and demonstrated that it is generally well below the standard half-the-wavelength  estimate. Our results also apply to other methods that rely on wave propagation and scattering  -- such as e.g. geophysical and ultrasound imaging.

\subsection*{Acknowledgments}
This work was partially supported by the Gordon and Betty Moore Foundation.


\begin{thebibliography}{99}
\bibitem{footnote}
{While the concept of error-free information recovery in the presence of noise may sound surprising, the Reader just experimentally tested it downloading this article (error-free)  over a noisy internet connection}.
\bibitem{footnote2}
For a given illumination field, each point scatterer can be treated as an effective point source.
\end{thebibliography}


%



\appendix

\section{Imaging model.} In its most general setting, the problem of (optical) imaging is essentially
the reconstruction of the object profile from scattering data. The formation of the desired image of the target
can be achieved using ``analog'' or ``digital'' tools, from lenses and projection screens in the former case and computational reconstruction of the object pattern on a computer screen in the latter. If the structure of the object is represented by its dielectric permittivity profile $\epsilon\left({\bf r}\right)$, the scattered electric field at the given freqeuncy $\omega$ is defined by the vectorial Lippmann-Schwinger equation
\begin{eqnarray}
{\bf E}\left({\bf r}\right) & = & {\bf E}_0\left({\bf r}\right) \nonumber \\
& + & k_0^2 \int d{\bf r'}  {\bf G}_0\left(k_0 \left| {\bf r} - {\bf r'}\right|\right)  \Delta\epsilon\left({\bf r'}\right) {\bf E}\left({\bf r'}\right), \ \ \ 
\label{eq:LSw} 
\end{eqnarray}
where ${\bf G}_0\left(k_0 \left| {\bf r} - {\bf r'}\right|\right)$ is the (dyadic) Green function for the medium surrounding the object, $\Delta\epsilon\left({\bf r}\right) \equiv \epsilon\left({\bf r}\right) - \epsilon_0$ is the difference between the permittivities of the object and of the surrounding medium, and $k_0 \equiv \sqrt{\epsilon_0} \ \omega/c$. 

Alternatively, the object may be represented as a collection of small (subwavelength) particles, with the individual (tensor) polarizabilities ${\bf \alpha}_i$, leading to
\begin{eqnarray}
{\bf E}\left({\bf r}\right) & = & {\bf E}_0\left({\bf r}\right) + 4 \pi k_0^2 \sum_i  {\bf G}_0\left(k_0 \left| {\bf r} - {\bf r}_i\right|\right) {\bf \alpha}_i  {\bf E}\left({\bf r}_i\right). \ \ \ \ \ \ \
\label{eq:LSwA} 
\end{eqnarray}
Note that these two formulations are essentially equivalent, as arbitrary dielectric permittivity profile can be expressed in terms of the electromagnetic response of a large group of small particles.\cite{Draine1994}

While Eqns. (\ref{eq:LSw}) and (\ref{eq:LSwA})  are linear in the electrical field, when treated as  inverse problems for the reconstruction of the unknown profile  
$\Delta\epsilon\left({\bf r}\right)$ and the distribution $\alpha_i$ from the given illumination field ${\bf E}_0\left({\bf r}\right) $ and the scattering data for ${\bf E}\left({\bf r}\right)$, they are essentially nonlinear in $\Delta\epsilon$ and $\alpha$.\cite{ChewBook} Physically, this nonlinearity originates from the multiple scattering effects within the object,
\cite{CuiChewYin2004} when  the actual field acting on the given object,  ${\bf E}\left({\bf r}\right)$, in addition to the incident field ${\bf E}_0$, also includes the contributions from the ``secondary'' waves scattered by the other parts of the object. While these multiple scattering  corrections can be substantial in acoustic and microwave  scattering,\cite{CuiChewYin2004}  for optical imaging of low-contrast media these are generally small.\cite{BornWolf} Note however that, when substantially present, these ``secondary'' waves due to multiple light scattering, can have a profound effect on the imaging resolution \cite{CuiChewYin2004} -- as the subwavelength structure of the object now functions as a high spatial frequency grating forming an effective structured illumination pattern.

In the language of scattering theory, the conventional optical imaging and microscopy corresponds to the limit of weakly scattering semi-transparent objects, that neglects multiple scattering contributions. The resulting first order 
Born approximation \cite{BornWolf} reduces the acting field in the integral of Eqn. (\ref{eq:LSw}) and the sum of Eqn. (\ref{eq:LSwA}) to the (a priori known) illumination field ${\bf E}_0$, thus leading to a linear inverse problem.
 
The resulting expressions can be further simplified in the radiation zone, when the detectors are placed in the far-field of the object, $k_0 \left| {\bf r} - {\bf r}_i\right| \gg 1$, thus reducing e.g. Eqn. (\ref{eq:LSwA}) to
\begin{eqnarray}
{\bf E}\left({\bf r}\right) & = &k_0^2 \sum_i \frac{ \left[ \left( {\bf r - r}_i \right)\times \alpha_i {\bf E}_0\left({\bf r}_i\right)\right] \times \left( {\bf r - r}_i \right)}{\left|  {\bf r - r}_i \right|^2} \nonumber \\
& \times & \exp\left( i k_0 \left|  {\bf r - r}_i \right| \right).
\label{eq:E1}
\end{eqnarray}
When the distance to the detector $r$ is much larger than the aperture $L$, $r \gg L$, for the far-field signal detected  in the given polarization and the wavevector ${\bf k}$ (see Fig. \ref{fig:schematics})
we find
\begin{eqnarray}
s\left({\bf k}\right)  & = & \sum_i  \alpha_i E_0\left({\bf \boldsymbol\rho}_i\right) \exp\left(i {\bf k}\cdot \boldsymbol\rho_i\right) + n\left({\bf k}\right), \ \ \ 
\label{eq:s1}
\end{eqnarray}
where ${\bf \boldsymbol\rho}_i \equiv \left(x_i, y_i\right)$,  ${\bf k} \equiv \left(k_x, k_y\right)$ with the magnitude $\left| {\bf k} \right| < k_0$, and $n$ is the noise in the detector positioned in the corresponding detector (see Fig. \ref{fig:schematics}).

Similarly, if the target is represented with the 2D permittivity mask $\epsilon(x,y)$ (corresponding to the 
Motti projection \cite{Motti}  of the actual 3D permittivity of the object), we obtain
\begin{eqnarray}
s\left({\bf k}\right)  & = & \int d^2{\bf \boldsymbol\rho} \ \Delta\epsilon\left({\bf \boldsymbol\rho}\right)  \ E_0\left({\bf \boldsymbol\rho}_i\right) \exp\left(i {\bf k}\cdot \boldsymbol\rho_i\right) + n\left({\bf k}\right). \ \ \ \ \ \ \ 
\label{eq:s2}
\end{eqnarray}

\section{Information entropy} 

The entropy $H\left[s\right]$ offers a the measure of the information received by the detector that returns the value of $s$, and is a functional of the statistical distribution of $s$:
\begin{eqnarray}
H\left[s\right] & = & - \int ds \ p\left(s\right) \ \log_2p\left(s\right),
\label{eq:hs}
\end{eqnarray}
When $s$ represents the scattered field detected in the imaging system, it's defined by the object structure and the illumination field profile. However, even in the absence of any stray light in the system, all detectors are
inherently noisy. As a result, for a given detected signal, there will always some uncertainty. This uncertainly is represented by the conditional information entropy $H\left[s \left|\right. o \right]$ of the detected signal for a given object, in terms of the conditional distribution $p\left(s \left|\right. o\right)$: 
\begin{eqnarray}
H\left[ s \left|\right. o\right] & = & - \int ds \ p\left(s \left|\right. o\right) \ \log_2p\left(s \left|\right. o\right).
\label{eq:hsc}
\end{eqnarray}
According to the Shannon's fundamental result,\cite{Shannon} the resulting information about the object is then given by the mutual information
\begin{eqnarray}
T & = & H\left[s\right] - H\left[ s \left|\right. o\right].
\label{eq:TA}
\end{eqnarray}

When the imaging system measures the continuous spectrum $s\left({\bf k}\right)$, the relevant entropies are defined by the functional integral
\begin{eqnarray}
H & = & - \int {\cal D}s\left({\bf k}\right)  \ p \ \log_2p,
\end{eqnarray}
where $p \equiv p\left[s\left({\bf k}\right)\right]$ for the entropy $H\left[s\right]$, and $p \equiv p\left[s\left({\bf k}\right)\right]$ for the entropy  $H\left[s\left|\right. o \right]$, while the functional integral is defined in the standard way
\begin{eqnarray}
\int{\cal D}\xi\left({k}\right) \equiv \lim_{M \to \infty} c_M \left[\Pi_{m=1}^M \int d\xi\left(k_m\right)\right],
\end{eqnarray}
where $c_M$ is the normalization constant.

\section{Mutual Information} 

The mutual information $T$ is defined \cite{Shannon} as the difference between the information entropy at the ``output'' 
$s\left({\boldsymbol k}\right)$ for the unconstraint ``input'', $H\left[\left\{ s\right\}\right]$, and the information entropy $H\left[\left\{ s\right\} \left|\right. E_0 \alpha\right]$ of the outpuf for fixed input $E_0\left({\boldsymbol \rho}\right) \alpha\left({\boldsymbol \rho}\right)$ -- see Eqn. (\ref{eq:s1}). For additive noise, the latter is simply equal to the noise entropy,
\begin{eqnarray}
H\left[\left\{ s\right\} \left|\right. E_0 \alpha \right] & = &  - \int {\cal D}n\left({\boldsymbol k}\right) \ P_n\left[ n\left({\boldsymbol k}\right)\right] \nonumber \\
& \times &  \log_2P_n\left[ n\left({\boldsymbol k}\right)\right], 
\label{eq:H_unconditional}
\end{eqnarray}
 where $P_n\left[n\left({\boldsymbol k}\right)\right]$ is the noise distribution function, and reduces to
 \begin{eqnarray}
H\left[\left\{ s\right\} \left|\right. E_0 \alpha \right] & = &  \int d{\bf k} \ \log_2\left[\pi e \sqrt{\langle \left| n\left({\bf k}\right)\right|^2 \rangle }\  \right]
\label{eq:Hnoise}
\end{eqnarray}
for uncorrelated Gaussian noise.

The unconditional output distribution, $P_n\left[ s\left({\boldsymbol k}\right)\right]$, is defined by both the noise and the target profile distribution $P\left[\left\{\alpha\right\}\right]$. While the latter does not necessarily reduce to a simple functional form, every single output component $s\left({\boldsymbol k}\right)$ corresponds to a sum of many such random variables -- see Eqn. (\ref{eq:s1}).  The central limit theorem then implies that the ``output'' statistics of 
$s\left({\boldsymbol k}\right)$ is described by the correlated multivariate normal distribution. Changing the path integral variables in (\ref{eq:H_unconditional}) using the orthogonal transformation that diagonalizes the corresponding covariance matrix, we obtain
\begin{eqnarray}
H\left[\left\{ s\right\} \right] & = & \sum_\lambda \ \log_2\left[\pi e \left(
2 \lambda \langle \left| n\left(k\right)\right|^2 \rangle \langle \left| \alpha -  \langle\alpha\rangle\right|^2\rangle
  \right. \right. \nonumber \\
& + & \left.\left. 
\lambda^2 \eta  \langle \left| \alpha -  \langle\alpha\rangle\right|^2\rangle^2 
+
 \langle \left| n\left(k\right)\right|^2 \rangle^2
 \right)^{1/2}  \right], \ \ \ \ \ 
\label{eq:Hlambda}
\end{eqnarray}
where $\lambda$'s are the eigenvalues of the discrete prolate spheroidal Slepian matrix \cite{Slepian1978}  
$
S_{{\boldsymbol k}_1 {\boldsymbol k}_2}  =  S\left({\boldsymbol k}_1 - {\boldsymbol k}_2\right),
$
with $\left| {\bf k}\right| < k_0$, where for a one-dimensional target,
$
S_1\left(q\right)  = {\sin\frac{qL}{2}}/{\sin\frac{q \Delta}{2}},
$
while for a rectanglular (square) aperture
$
S_2\left({\bf q}\right)  =  S_1\left(q_x\right) S_1\left(q_y\right).
$
The eigenvalue spectrum of the Slepian matrix has the characteristic step shape, showing significant $k_0L$ significant 
eigenvalues ($\lambda \approx  L / \Delta$) and remaining  insignificant eigenvalues ($\lambda \approx 0$)  separated by a narrow transition band.\cite{Landau1965,SlepianSonnenblick1965} The eigenvalue sum in Eqn. (\ref{eq:Hlambda}) can be therefore calculated analytically, which together with Eqns. (\ref{eq:Hnoise}) and (\ref{eq:T}) yields (\ref{eq:delta}).

\section{Resolution Limit for Structured Illumination} 

In the case of structured illumination, for the resolution limit we obtain
\begin{eqnarray}
\delta^{SE} & = &   \frac{\lambda}{4} \ \frac{1}{\log_2\sqrt{ 1 + 2 \ {\rm SNR} + \eta \ {\rm SNR}^2  } + {\cal{O}}\left(1/k_0 L\right)},  \ \ \ \ \ \ \  
\\ 
\delta_M^{SE} &  = &  \frac{\lambda}{4} \ \frac{\log_2M}{\log_2\sqrt{ 1 + 2 \ {\rm SNR} + \eta \ {\rm SNR}^2  } + {\cal{O}}\left(1/k_0 L\right)}. \ \ \ \ \ \ \ 
\label{eq:deltaSE}
\end{eqnarray}


\end{document}